\documentclass[a4paper]{jpconf}

\usepackage{graphics}
\usepackage{graphicx}
\usepackage{bm}
\usepackage{amsmath}
\usepackage{amsfonts}
\usepackage{amssymb}
\usepackage{latexsym}
\usepackage{color}

\definecolor{darkred}{rgb}{.8,0,0}

\definecolor{darkblue}{rgb}{0,0,.8}

\newcommand{\be}{\begin{equation}}\newcommand{\ee}{\end{equation}}
\newcommand{\bea}{\begin{eqnarray}}\newcommand{\eea}{\end{eqnarray}}
\newcommand{\brr}{\begin{array}}\newcommand{\err}{\end{array}}
\newcommand{\bit}{\begin{itemize}}\newcommand{\eit}{\end{itemize}}
\newcommand{\ben}{\begin{enumerate}}\newcommand{\een}{\end{enumerate}}

\def\prl{{\it Phys. Rev. Lett.}\ }

\def\apj{{\it Ap. J.}\ }

\def\apj{{\it Ap. J.}\ }

\def\lan{\langle}
\def\lf{\left}

\def\non{\nonumber}
\def\ran{\rangle}

\def\ri{\right}

\def\si{\sigma}
\def\om{\omega}

\def\1{{_{1}}}\def\2{{_{2}}}

\begin{document}

\title{Quantum field theory of axion-photon mixing and vacuum polarization}

\author{Antonio  Capolupo }
\address{ Dipartimento di Fisica E.R.Caianiello and INFN Gruppo Collegato di Salerno,
  Universit\'a di Salerno, Fisciano (SA) - 84084, Italy}

\begin{abstract}
We report on  recent results obtained by analyzing axion--photon mixing in the framework of quantum field theory.
We obtain corrections to the oscillation formulae and we  reveal a new effect of the vacuum polarization due to the non-zero value of the vacuum energy for the component of the photon polarization mixed with the axion. The study of axion--photon mixing in curved space is also presented. Numerical analysis show that some  quantum field theory effect of axion--photon mixing, in principle, could be detected experimentally.

\end{abstract}

\section{Introduction}

Axions are light neutral particles with mass $10^{-6} - 10^{-2}$ eV  \cite{peccei,peccei1}
that, together  with ultra-light axions (ULAs) with masses   of $\sim 10^{-22}$ eV \cite{Witten1984}--\cite{Hui2017} are considered as possible candidates for dark matter \cite{Khmeinitsky:2013lxt}--\cite{Emami2018}.
An interesting property of axions and axion-like particles (ALPs) is axion--photon mixing and oscillation in the presence of strong magnetic fields \cite{raffelt}.  This phenomenon (which is negligible for ULAs)  could be used to detect axions with masses  of $10^{-3}$ and $10^{-2}$ eV. However, at the moment there is no experimental demonstration  of the existence of axions   \cite{PVLAS1}--\cite{17}.
Therefore the study of possible experimental setups that
could reveal axions and ULAs, and the complete theoretical understandings of the axions and  of axion--photon mixing are needed.

 In this paper we report the results presented in Ref.\cite{CapolupoAssioni}, obtained by analyzing axion--photon mixing in the framework of quantum field theory (QFT). 
 We present new oscillation formulae for the axion--photon system and demonstrate the presence of QFT vacuum polarization induced by the condensate structure of the vacuum for mixed particles.
We show that the QFT corrections to the amplitudes of the oscillation formulae, which are negligible for neutrinos and kaons, are detectable in principle in the case of axion--photon mixing. We also study axion--photon mixing in curved space and we provide numerical estimates of the energy density of the vacuum for axions and photons.

It should be noted  that axion--photon mixing, at the difference of the mixing of neutrinos and bosons such as  kaons, is characterized by the mixing of particles with different nature. Moreover, only one component of the photon polarization state couples and oscillates with the axion, generating a vacuum polarization that is absent in the other mixed systems. Therefore the  results achieved by the QFT analysis of
 the mixing of neutrinos and mesons \cite{Blasone:1998hf}--\cite{Blasone:2004yh-1} cannot be applied immediately to axion--photon mixing and 
 specific adjustments are required.

The  paper is organized as follows. In Section 2, we present the QFT treatment of axion--photon mixing and we show new oscillation formulae.
In Section 3, we analyze axion--photon mixing in curved space--time and we show a new polarization effect of the vacuum and in Section 4 we give our conclusions.

\section{Axion--photon mixing}

By neglecting the Heisenberg--Euler term due to loop correction in QED \cite{Heisenberg}-\cite{Dobrich}, $\frac{\alpha^{2}}{90 m_{e}^{4}}\lf[(F_{\mu \nu } F^{\mu \nu })^{2}
+ \frac{7}{4}(F_{\mu \nu } \tilde{F}^{\mu \nu })^{2}\ri]$, the Lagrangian density describing the  ALP--photon system is
\bea\label{1}
L = - \frac{1}{4} F_{\mu \nu} F^{\mu \nu} + \frac{1}{2} (\partial_{\mu} a \partial^{\mu} a - m_{a}^{2} a^{2})
 + \dfrac{g_{a \gamma \gamma}}{4} \, a \, F_{\mu \nu}\tilde{F}^{\mu \nu}\,.
\eea
Here the first two terms are the Lagrangian densities of the free photon and axion, respectively,
and the last term is the interaction of two photons with the axion pseudoscalar field $a$ in the presence of a magnetic field, which is responsible for axion--photon mixing \cite{PDG}.
In Eq.(\ref{1}), $\tilde{F}_{\mu \nu } = \frac{1}{2}\epsilon_{\mu \nu \rho \sigma}F^{\rho \sigma} $ is the dual electromagnetic tensor and $g_{a\gamma \gamma}\equiv g \equiv g_{\gamma} \alpha/ \pi f_{a} $ is the axion--photon coupling, with   $g_{\gamma} \sim 1$, $\alpha =1/137$ and $f_{a}$   decay constant for ALPs.

We consider a laser  beam propagating through a magnetic field $\textbf{B}$  in the direction perpendicular to $\textbf{B}$.
In this case, the photon polarization state $  \gamma_{\bot}$ perpendicular to $\textbf{B}$ and to the direction of propagation of the beam decouples.
 Then, the mixing characterizes only the axion and  the photon polarization state $\gamma_{\|}$ parallel to the magnetic field and the propagation equations can be written as:
$
\left(\omega - i \partial_{z} + \textit{M}  \right)\left(
                                                \begin{array}{c}
                                                  \gamma_{\|} \\
                                                  a \\
                                                \end{array}
                                              \right) = 0\,,
$
where, $\textit{M}$ is the mixing matrix
$
\textit{M} = -\frac{1}{2\omega}\, \left(
               \begin{array}{cc}
                 \omega_{P}^{2} & - g \omega B_{T} \\
                 - g \omega B_{T} & m_{a}^{2} \\
               \end{array}
             \right).
$
Here,   the magnetic field $\textbf{B}$ coincides with the purely transverse field $ {B}_{T}$, and $\omega_{P} = (4 \pi \alpha N_{e} /m_{e} )^{1/2}$ is the plasma frequency that is equal to zero in the case of propagation in the vacuum. Moreover, $N_{e}$ is the electron density and $m_{a}$ is the axion mass.
 The matrix \textit{M} can be diagonalized by means of a rotation
\bea\label{mixing}
\left(
  \begin{array}{c}
    \gamma^{\prime}_{\|}(z) \\
    a^{\prime}(z) \\
  \end{array}
\right) = \left(
            \begin{array}{cc}
              \cos \theta & \sin \theta \\
              -\sin \theta & \cos \theta \\
            \end{array}
          \right)
          \left(
            \begin{array}{c}
              \gamma_{\|}(z) \\
              a(z) \\
            \end{array}
          \right)\,,
\eea
where $\gamma^{\prime}_{\|} $ and $a^{\prime}$ are the ``free''
fields with definite masses, $\gamma_{\|} $ and $a$  are
the fields of the mixed particles   and
 $\theta = \frac{1}{2} \arctan \displaystyle{\left(\frac{2 g \omega B_{T}}{m_{a}^{2}-\omega^2_P} \right)}$ is the mixing angle.
Since  any polarization component of a neutral vectorial field  can be represented by a neutral scalar field,
then the decoupling of the photon polarization state $  \gamma_{\bot}$ allows us to treat  $\gamma_{\|} $ as a neutral scalar field.
Thus axion--photon mixing can be described by
the formalism presented in Ref.\cite{CapolupoPLB2004} used for the mixing of neutral particles in the QFT framework.

We invert Eqs.(\ref{mixing}), we quantize $\gamma^{\prime}_{\|}(x)$ and $a^{\prime}(x)$ in the usual way and we
express the mixed fields   by means of the mixing generator $G_\vartheta(t)$  as
$
\gamma_{\|} (z)  = G^{-1}_\vartheta(t)\; \gamma_{\|}^{\prime} (z)\; G_\vartheta(t)
$
and
$
a(z) = G^{-1}_\vartheta(t)\; a^{\prime}(z)\; G_\vartheta(t)\,,
$
where $\vartheta = - \theta$. In a similar way, we can define the
annihilators
  for $\gamma_{\|} $ and $a$  as $\alpha_{{\bf k},\gamma_{\|}}(\vartheta ,t) \equiv G^{-1}_\vartheta(t) \;
\alpha_{{\bf k},\gamma_{\|}^{\prime}}(t)\;G_\vartheta(t)$ and $\alpha_{{\bf k},a}(\vartheta ,t) \equiv G^{-1}_\vartheta(t) \;
\alpha_{{\bf k},a^{\prime}}(t)\;G_\vartheta(t)$,
 such that $\alpha_{{\bf k},\gamma_{\|}}(\vartheta ,t)
|0(\vartheta, t) \ran_{\gamma_{\|}, a} = \alpha_{{\bf k},a}(\vartheta ,t)
|0(\vartheta, t) \ran_{\gamma_{\|}, a} = 0$.
Here $
  |0(\vartheta, t) \ran_{\gamma_{||}, a} \equiv
G^{-1}_ \vartheta(t)\; |0 \ran_{ \gamma_{||}^{\prime},a^{\prime}}\,
$
is the vacuum for mixed fields which in the infinite volume limit is orthogonal to the vacuum for free fields $|0  \ran_{\gamma^{\prime}_{||}, a^{\prime}}$. Moreover,
$\alpha_{{\bf k},\gamma_{\|}^{\prime}}(t) = \alpha_{{\bf k},\gamma_{\|}^{\prime}}\, e^{-i \om_{k, \gamma_{\|}^{\prime}} t} $,
$ \alpha_{{ \bf k},a^{\prime}}(t) = \alpha _{{ \bf k},a^{\prime}}\, e^{i \om_{k, a^{\prime}} t} $, are the annihilators
 for the free fields $\gamma_{\|}^{\prime}  $ and $a^{\prime}$, respectively, where $
\omega_{k, \gamma_{\|}^{\prime}} = \omega_k + \Delta_{+}\,,
$
and
$
\omega_{k, a^{\prime}} = \omega_k + \Delta_{-}\,,
$
 with $\omega_k$ photon energy
and
$
\Delta_{\pm} = - \frac{\omega_{P}^{2}+ m_{a}^{2}}{4 \omega_k} \pm \frac{1}{4 \omega_k} \sqrt{(\omega_{P}^{2}- m_{a}^{2})^{2} + (2 g \omega_k B_{T})^2}\,.
$
Explicitly, we have:
$
\alpha_{{\bf k},\gamma_{\|}}(t) = \cos\vartheta  \alpha_{{\bf k},\gamma_{\|}^{\prime}}(t)
+ \sin\vartheta
\lf( |\Sigma_{{\bf k}}|  \alpha_{{\bf k},a^{\prime}}(t) +
|\Upsilon_{{\bf k}}| \alpha^{\dag}_{{\bf k},a^{\prime}}(t) \ri)\, ,
$
and  similar for
$
\alpha_{{\bf k},a}(t)
$,
where the Bogoliubov coefficients are
 $|\Upsilon_{{\bf k}}|\equiv  \frac{1}{2} \lf(
\sqrt{\frac{\om_{k,\gamma_{||}^{\prime}}}{\om_{k,a^{\prime}}}}
- \sqrt{\frac{\om_{k,a^{\prime}}}{\om_{k,\gamma_{||}^{\prime}}}} \ri)\,,
$
  and
$|\Sigma_{{\bf k}}|  = \sqrt{1 + |\Upsilon_{{\bf k}}|^{2}}\,. $

Notice that, the vacuum for axions and photons is given by the following product:
  \bea
  |0(\vartheta,t)\rangle_{\gamma ,a} = |0(\vartheta,t)\rangle_{\gamma_{||},a} \otimes
  |0\rangle_{\gamma^{\prime}_{\bot}},
  \eea
 where   $|0(\vartheta, t) \ran_{\gamma_{||}, a}$ is the vacuum for  $\gamma_{||}$ and $a$,
 and $|0\rangle_{\gamma^{\prime}_{\bot}}$ is the vacuum for the unmixed component $ {\gamma^{\prime}_{\bot}}$.
 The state $|0(\vartheta, t) \ran_{\gamma_{||}, a}$
   has a  structure of
 condensed particles, with the condensation density   given by:
$
\,_{\gamma_{||}, a}\lan0(\vartheta, t) |\alpha_{{\bf k},\gamma_{||}^{\prime}}^{\dag} \alpha_{{\bf k},\gamma_{||}^{\prime}} |0(\vartheta, t) \ran_{\gamma_{||}, a} = \sin^{2} \vartheta |\Upsilon_{{\bf k}}|^{2}\,.
 $ By contrast,  $|0\rangle_{\gamma^{\prime}_{\bot}}$ has a trivial structure and it does not exhibit a condensed structure. The presence of a condensate only for the component $\gamma_{||}$  produces the polarization of the vacuum for axions and photons.

The QFT oscillation formulae for the axion--photon system are given by the expectation value of the momentum operator for mixed fields ${\cal P}_\si(t)  = \int d^3 x  \,\pi_{\sigma}(x)\,\nabla\,\phi_{\sigma}(x)$ at $t\neq 0$ on the initial state of the photon   $|  \alpha_{{\bf k},\gamma_{\|}}\ran = \alpha^{\dag}_{{\bf k},\gamma_{\|}}(0) |0(\vartheta, 0) \ran_{\gamma_{\|}, a}\,$,
normalized to its initial value:
 $
 {{\cal P}}^{\bf k}_{\gamma \rightarrow \si}(t) \equiv  \frac{  \lan {
\alpha}_{{\bf k},\gamma_{\|}} | {\cal P}_\si(t) | {\alpha}_{{\bf k},\gamma_{\|}}\ran
\,}{  \lan { \alpha}_{{\bf k},\gamma_{\|}} | {\cal P}_\si(0) | {\alpha}_{{\bf
k},\gamma_{\|}}\ran  , }
 $ with $ \si = \gamma_{\|}, a.$
  Explicitly, we have:
  \bea
\label{Amomentum}
 {\cal P}^{\bf k}_{\gamma \rightarrow \gamma }(t) &=&
1 - \sin^{2}( 2 \theta) \Big[ |\Sigma_{{\bf k}}|^{2} \; \sin^{2}
\lf( \Omega_{k}^{-} t \ri)
- |\Upsilon_{{\bf k}}|^{2} \; \sin^{2} \lf( \Omega_{k}^{+} t
\ri) \Big] \, ,
\\[4mm]
{\cal P}^{\bf k}_{ \gamma \rightarrow a}(t)&=&
\sin^{2}( 2 \theta)\Big[ |\Sigma_{{\bf k}}|^{2} \; \sin^{2} \lf(
\Omega_{k}^{-} t \ri)
- |\Upsilon_{{\bf k}}|^{2} \;
\sin^{2} \lf( \Omega_{k}^{+} t \ri) \Big]  ,
\label{Bmomentum}
\eea
where $\Omega_{k}^{\pm}=\frac{\omega_{k,a^{\prime}} \pm \omega_{k,\gamma_{\|}^{\prime}}}{2}$.
\begin{figure}[t]\centering
\begin{picture}(300,180)(0,0)
\put(10,20){\resizebox{8.0 cm}{!}{\includegraphics{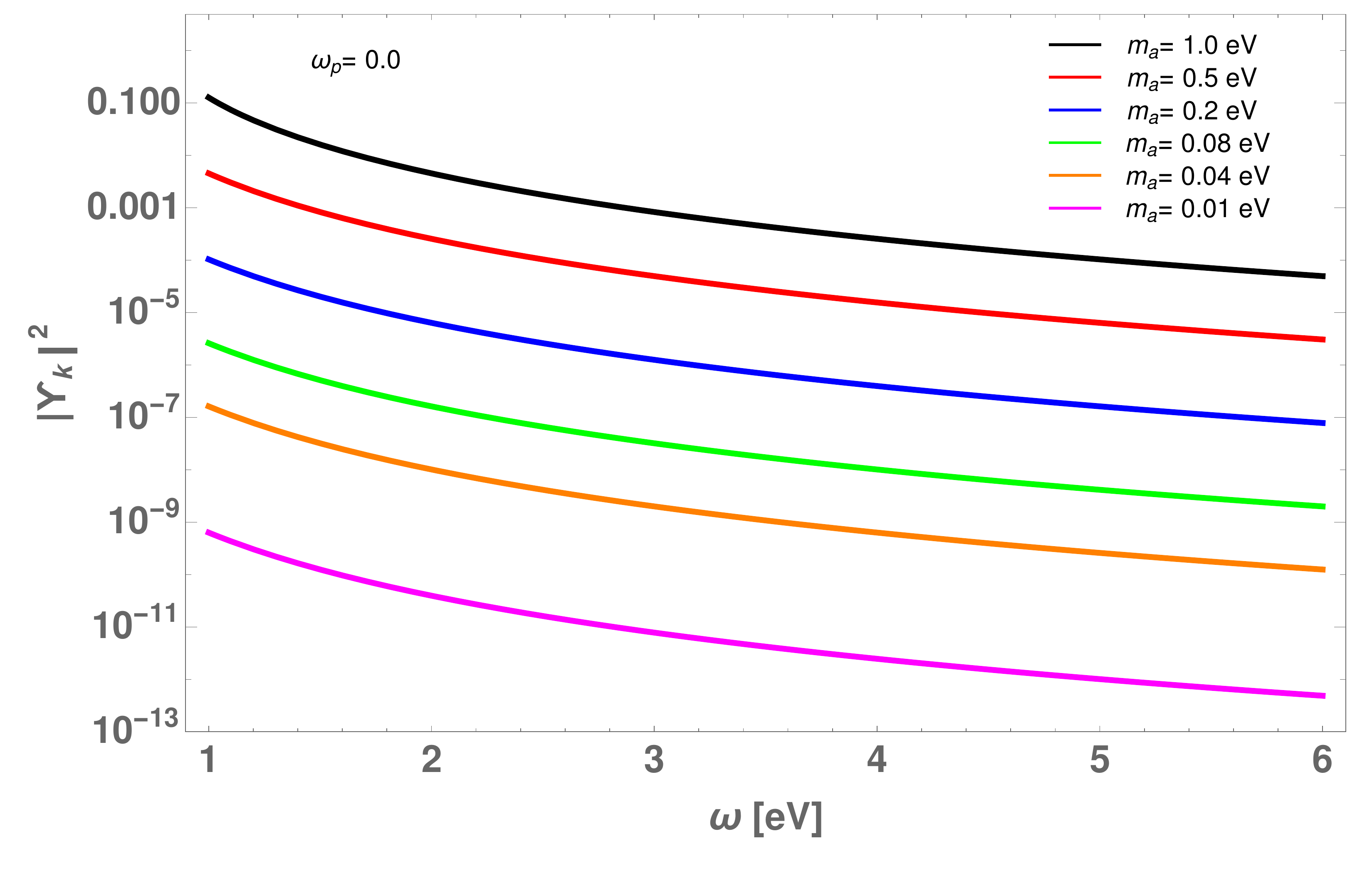}}}
\end{picture}\vspace{-1cm}
\caption{\em (Color online) Plots of $|\Upsilon_{\bf k}|^2$ as a function of the photon energy for $\omega_p = 0$.}
\label{pdf}
\end{figure}
In Fig.1 we plot   the coefficient $|\Upsilon_{\bf k}|^2$ which multiplies the high-frequency oscillation term in Eqs. (\ref{Amomentum}) and (\ref{Bmomentum}). We assume a photon energy  $\omega \simeq 1.5 eV$, a coupling constant   $g_{a \gamma} \sim 1.4 \times 10^{-11} GeV^{-1}$ \cite{giannotti} and an axion mass   $m_{a} \in [1 - 10^{-2}] eV$.
The plots show that values of $|\Upsilon_{\bf k}|^2$ which are negligible in the case of very low axion masses,  in principle, are detectable in the case of ``heavy'' axions, (axions with masses up to $0.2 eV$ are expected in some models  \cite{DiLuzio}).
The values of $|\Upsilon_{\bf k}|^2$ are
much higher than the corresponding field-theoretical corrections for the other mixed systems such as kaons and neutrinos  \cite{CapolupoPLB2004}.
Therefore, in principle, the axion--photon system could allow to test the QFT effects.
Finally, we note that the presence of a gas in the conversion region increases the oscillation rate   because the gas induces the effective mass of photon  $m_\gamma = \frac{\hbar}{c^{2}}\omega_{P}$. However, the plasma presence reduces the value of $|\Upsilon_{\bf k}|^2$, which in any case remain   detectable \cite{CapolupoAssioni}.

\section{  Axion--photon mixing in curved space--time  and vacuum polarization}

We analyze axion--photon mixing in curved space-time, by considering a homogeneous magnetic field which coincides with its transverse component, $B = B_T$ and we study the energy density and the pressure of   the vacuum for mixed fields.

 The unmixed scalar fields $\psi^{C}_{i} = \gamma^{\prime C}$, $a^{\prime C}$ in curved space-time are
\bea\label{psi}
\psi^{C}_{i}({\bf x},t) = J^{-1}({\bf x},t)\, \psi_{i}({\bf x},t)\, J({\bf x},t)
= \frac{1}{\sqrt{V}}\sum_{\bf k} \lf[u_{{\bf k},i}({\bf x},t)\alpha_{{\bf k},i} + u^{*}_{{\bf k},i}({\bf x},t)\alpha^{\dag }_{{\bf k},i}  \ri],
\eea
where $J$ is the generator of the curvature (which depends on the metric considered), $\psi_{i} = \gamma^{\prime  }$, $a^{\prime  }$ are the free fields operator in flat space--time, and $u_{{\bf k},i}({\bf x},t) = \zeta_{{\bf k},i}( t) e^{i {\bf k}\cdot{\bf x}}$ are the mode functions that can be expressed  analytically only in particular cases.
  The vacuum state for $\psi^{C}_{i}$ is   $|0 ({\bf x},t)\ran^{C}_{ \gamma_{\|}^{ \prime},a^{  \prime}} = J^{-1}({\bf x},t) |0 \ran_{ \gamma_{\|}^{\prime},a^{\prime}}\,,$ and the curved mixed vacuum is
 $|0(\vartheta, x) \ran^{C}_{\gamma_{\|}, a} =  (G^{C}_ \vartheta(x))^{ -1}\; J^{-1}(x) |0 \ran_{ \gamma_{\|}^{\prime},a^{\prime}}\,,$
where $G^{C }_ \vartheta(x)$ is the mixing generator for fields in curved space-time \cite{CapolupoAssioni}.
We take into account that for any curved space, the covariant derivative for scalar fields is equal to the ordinary derivative,
 and we compute the expectation value of the energy momentum tensor  density $T_{\mu\nu}(x)$ for $\gamma_{\|}^{\prime}, a^{\prime}$
 on the   mixed vacuum $|0(\vartheta, x) \ran^{C}_{\gamma_{\|}, a}  ,$
i.e.,
$
\Xi^{\|}_{\mu \nu} \,  \equiv \,  ^{C}_{\gamma_{\|}, a} \langle 0 ( \vartheta, x)|: T _{\mu \nu}(x): | 0 (\vartheta, x)\rangle^{C}_{\gamma_{\|}, a}\,.
$
Here $:...:$ indicates   the normal ordering with respect to  $|0 ({\bf x},t)\ran^{C}_{ \gamma_{\|}^{ \prime},a^{  \prime}}$.
 As shown in Ref.\cite{CapolupoAssioni}, the off-diagonal components of $\Xi^{\|} _{\mu \nu}(x)$ are zero for a homogeneous and isotropic universe, as well as for diagonal metrics, thus the energy density and pressure  are
 $
\rho_{\|}  = g^{00}\; \Xi^{\|}_{00} $,
and
$
p_{\|}  = - g^{j j}\; \Xi^{\|}_{ jj } $ (no summation on the index $j$ is intended), respectively.
Then, we obtain the  state equation of the cosmological constant: $p_{\|} = - \rho_{\|}$, where
 \bea\label{integral}
 \rho_{\|}  = \frac{\Delta m^{2}   \sin ^{2}\theta}{2 V}  \sum_{{\bf k}}   \lf(|\zeta_{{\bf k},1}( t)|^{2}  - |\zeta_{{\bf k},2}( t)|^{2} \ri),
 \eea
and $\Delta m^{2}= |m_{ a^{\prime}}^{2} - m_{\gamma^{\prime}}^{2}|$.

Astrophysical systems that produce strong magnetic fields are pulsar and neutron stars. In order to study of Eq.(\ref{integral}) for these object we should consider the Schwarzschild metric.
In this case, the explicit form of   $\rho_{\|}$ is difficult.
An estimate of $\rho_{\|}$, can be obtained by considering the
  Minkowski metric for simplicity. This means that we neglect the gravitational effects induced by the Schwarzschild  metric.
       This assumption is reasonable for many systems generating magnetic fields, because  their radius is very small.
       In the Minkowski metric, the explicit form of the energy density is:
    \bea\label{integral3}\non
 \rho_{\|} & = & \frac{\Delta m^{2}   \sin ^{2}\theta}{8 \pi^{2}} \Big\{  \lf( m_{a^{\prime}}^{2}- \sqrt{ m_{a^{\prime}}^{4}+ 4 g^{2}B_{T}^{2}K^{2}}
 \ri)
   -
 \lf( m_{a^{\prime}}^{2}+ g^{2}B_{T}^{2} \ri)
 \Big[\tanh^{-1} \lf(\frac{  m_{a^{\prime}}^{2}  }{ m_{a^{\prime}}^{2}+ g^{2}B_{T}^{2}} \ri)
\\
 & - &
  \tanh^{-1} \lf(\frac{\sqrt{ m_{a^{\prime}}^{4}+ 4 g^{2}B_{T}^{2}K^{2}}}{ m_{a^{\prime}}^{2}+ g^{2}B_{T}^{2}} \ri)
\Big ]\Big\},
 \eea
 where   $K$ is the cut-off on the momenta.
Since, for $x \in R$, the function $\tanh^{-1}(x)$ is defined in the domain:
$
-1 < x < 1
$,
  then Eq.(\ref{integral3}) imposes
 a condition on the cut-off $K$   given by:
$
-1 < \frac{\sqrt{ m_{a^{\prime}}^{4}+ 4 g^{2}B_{T}^{2}K^{2}}}{ m_{a^{\prime}}^{2}+ g^{2}B_{T}^{2}}< 1 ,
$
which implies,
$
K < \frac{\sqrt{2 m_{a^{\prime}}^{2} + g^{2}B_{T}^{2}}}{2}\,.
$

 We now give a numerical estimation of Eq.(\ref{integral3}) for astrophysical objects and for terrestrial experiments.

 {\it -  Astrophysical systems:}
  we consider the following values of the parameters:  $m_{a^{\prime}}\sim 2 \times 10^{-2}eV$, $g \sim 10^{10}GeV^{-1}$, $B_{T} \sim (10^{15} - 10^{16})G $ (which can be obtained for different astrophysical objects),   $\omega \sim 100 eV$,  $K\sim 10^{-2}eV$ (which satisfies the upper bound on $K$). We have: $\rho_{\|} \sim 10^{-47}GeV^{4}$, which is of the same order as the estimated value of the dark energy.
However, $\rho_{\|}$ does not contribute to the dark energy, because it appears only in the regions where the magnetic field is localized.
Moreover, in these regions other fields would have a larger influence on the energy momentum tensor.

{\it - Terrestrial experiments:}
 We consider  $B = 45 T$, $\omega \sim 10 eV$ and $g \sim 10^{10}GeV^{-1}$. For these values of the parameter, the term  $\tanh^{-1} \lf(\frac{  m_{a^{\prime}}^{2}  }{ m_{a^{\prime}}^{2}+ g^{2}B_{T}^{2}} \ri)$ in Eq.(\ref{integral3}), imposes that    $\rho_{\|}$ has a finite value only for values of axion mass less than $3 \times 10^{-7}eV.$
For example, for $m_{a^{\prime}}\sim 2 \times 10^{-7}eV$, we have $\rho \sim 10^{-66}GeV^{4}$.

It should be noted  that the vacuum energy density and pressure are non zero  only for the component parallel to the magnetic field of the photon polarization.  On the contrary, the contributions $\rho_{\bot}$ and $p_{\bot}$   of the component perpendicular to the magnetic field of the photon polarization $\gamma_{\bot}$ are equal to zero:
$
\Xi^{\bot}_{\mu \nu}(x)   \equiv    _{\gamma_{\bot} } \langle 0({\bf x},t)  |: T _{\mu \nu}(x): | 0({\bf x},t)  \rangle_{\gamma_{\bot} } = 0 .
$
This fact leads to a QFT polarization of the vacuum.
The QFT vacuum polarization here presented could be detected in next experiments on axion--photon mixing.

\section{Conclusions}

We analyzed axion--photon mixing in  the QFT framework.
We derived new oscillation formulae for axion--photon transitions and revealed a  vacuum polarization
due to the condensate structure of the vacuum for mixed fields. These effects  are not expected in previous studies of axion--photon mixing in the framework of quantum mechanics.
Our numerical analysis shows that, in principle, the QFT effects can be detected in laser beam experiments.

\section*{Acknowledgments}

I acknowledge partial financial support from MIUR and INFN and  the COST Action CA1511 Cosmology
and Astrophysics Network for Theoretical Advances and Training Actions (CANTATA)
supported by COST (European Cooperation in Science and Technology).

\medskip

\end{document}